\documentclass[12pt]{amsart}
\usepackage{geometry} % see geometry.pdf on how to lay out the page. There's lots.
\usepackage{graphicx}

\title[Multiscale carbonation]{Multiscale carbonation reactions: Status of things and two modeling exercises related to cultural heritage}
\author{Adrian Muntean}
\address{Department of Mathematics and Computer Science\\ Karlstad University, Sweden\\
email: adrian.muntean@kau.se}
\begin{document}
\begin{abstract}
Having in mind as target audience  beginner researchers working in the field of cultural heritage, we present succinctly the concept of two-scale modeling of reaction-diffusion problems as it fits to scenarios where the action of the carbonation reaction is relevant. We briefly review well-known contributions concerning multiscale concrete carbonation processes, and finally, we point out two related multiscale modeling exercises. 

The scope of these notes is twofold: Promoting the language of multiscale modeling, we invite the applied mathematician to pick some of the target problems from the context of cultural heritage. On the other hand, we invite the experimentalist to talk to the applied mathematician whenever the laboratory experiments are unable to answer questions for instance about the long time behavior of materials exposed to the ingress of various chemical species, humidity, and/or temperature (as it is the case of historical monuments and buildings in urban areas) because modeling and simulation approaches might provide at least provisory hints on what could happen further once measurements stop, or when the situation of interest takes place  actually outside the laboratory.
\vskip1cm

\end{abstract}
\maketitle
%\tableofcontents

\section{Introduction}

\subsection*{Motivation} The role of these notes is to contribute with transfer of multiscale mathematical modeling technology to researchers involved in those cultural heritage studies where changes into the microstructure of the target (to be preserved, or to be reconstructed) material is of interest. This text is an expression of own decision to contribute with research to the conservation of the memory of the past and is in the same spirit of the developments reported recently by the mathematical community; see, for instance, \cite{MACH21} and \cite{Carola} and references cited therein.

To fix ideas, we formulate a generic distributed-microstructure model for a structured diffusion transport taking place in porous media, involving the  carbonation reaction; compare section \ref{sec-abstract} for further details. Such a modeling approach overlaps partly with the classical framework of the so-called dual-porosity or double-porosity modeling, hence it is not new {\em per se},  but it seems to be unexplored in the cultural heritage context, that is why our interest. One may though wonder:
\begin{center}
What has carbonation reaction to do with cultural heritage preservation?
\end{center}
 As the authors of the article \cite{Michel} indicate,  "among building structures, concrete monuments, churches and houses are an important part of our cultural heritage and correspond to a period of architectural history which was promising and enthusiastic for the social development of humanity" ({\em loc. cit.}). To address the posed question,  we only wish to stress at this stage that, by itself, the carbonation reaction does not destroy the porous material. Instead, it lowers the local pH values making the material in question rather vulnerable what concerns the penetration of aggressive ions (like sulphates or chlorides). In this way, the carbonation reaction does affect directly the durability and service life of the respective structure, therefore its relevance.

\subsection*{Organization of the work}
 We describe in section \ref{sec-abstract} a generic two-scale reaction-diffusion system that is applicable to unsaturated porous media exposed to chemical species that have the ability to dissolve in the water phase of the medium and either react there with the existing active chemicals or return to the air phase of the medium. A brief literature review around this theme is the subject of section \ref{literature}, while section \ref{tasks} contains two multiscale modeling tasks (on purpose vaguely formulated). 
 
 We hope that these modeling tasks will act like  an appetizer for further multidisciplinary discussions around this subject and related topics. 

%\tableofcontents

\section{Mass balance equations for a generic two-scale  setting}\label{sec-abstract}

The presentation of the main modeling ideas outlined in this section is very much influenced by the excellent texts of R. E. Showalter
\cite{Show,Show0} as well as by his follow-up works in the same context (including the topic of fissured-media equations). They have influenced very much the further development of modeling with distributed microstructures in the context of the topic "reactive flow in porous media"; see, for instance, the list of references indicated at the end of this text. 
We take over the flavor of the presentation of the two-scale balance equations derived originally for parallel flows in structured porous media and apply it to a diffusion context. The main player is a population of $CO_2$ molecules perceived at the continuum level, hence in a deterministic framework. 

To be able to describe mathematically the basic physical scenarios we have in mind to be behind the carbonation process, we need to agree on a set of notations. 
At each $x\in\Omega$, there is a cell $Y(x)$ (local, microscopic), $y\in Y(x)$. Assuming that every pore of the porous media $\Omega$ has a wet part (typically, the wet part is a tiny layer of water, possibly connected,  clinging on the pore's fabric),  this cell corresponds to precisely such wet part. This is normally the place where the local chemistry takes place.  The assumption also requires that the volume of the cell is non-degenerate (i.e. $vol(Y(x))=|Y(x)|>0$), on the other hand it is not necessarily small\footnote{The smallness of the volume of the cell $|Y(x)|$ is likely to be needed if one plans to have a rigorous derivation of the two-scale models presented within this framework from the viewpoint of asymptotic periodic homogenization techniques; see e.g. the discussion in \cite{Show0} for the case of periodically-distributed microstructures $|Y(x)|=|Y|= const$; see also the original work \cite{ADH}. If the smallness assumption is not accounted for, then a potentially fast growth of the volume of the microstructures $Y(x,t)$ might become possible; see e.g. the very recent contribution \cite{Gabriella} where porosity variations are accounted for in the context of concrete carbonation. Consequently, all effective quantities (porosities, transport tensors, etc.) will become functions of $x,t$. Mind that the porosity $\phi$ is simply a number in $[0,1]$ when one treats the periodic case.  }. Here 
$t\geq 0$ is the  "common" time variable. The main unknowns are
\newline
\begin{center}
$u(x,t):=[CO_2(g)](x,t)$ -- the global/macroscopic mass concentration of $CO_2(g)$,
\end{center}
and respectively,
\begin{center}
$U(x,y,t):=[CO_2(aq)](x,y,t)$ -- the local/macroscopic mass concentration of $CO_2(aq).$
\end{center}

\begin{figure}
\includegraphics[width = 0.35\textwidth]{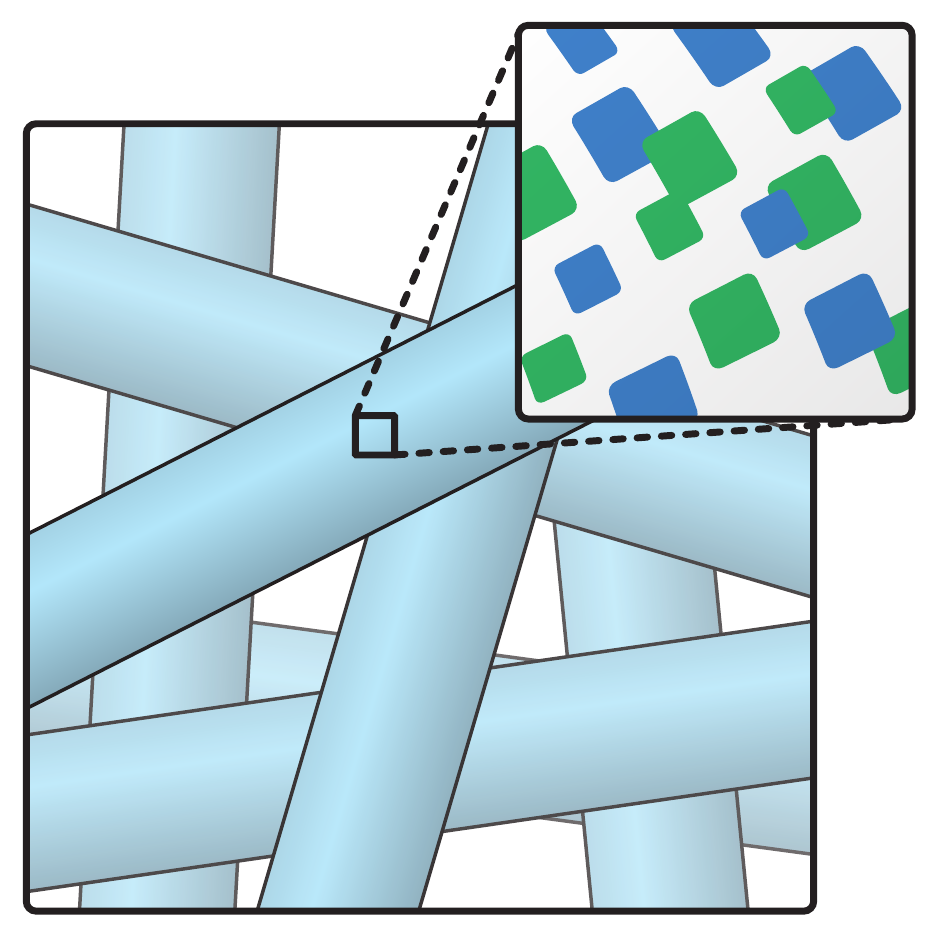}
\caption{Cartoon pointing out a macroscopic domain where in each point of the domain one has the possibility to zoom inside the microstructure, explore it, and then zoom out back to the macroscopic level.}
\end{figure}

The main task is now to write down the mass-balance equations for $u$ posed in the set $\Omega$, and respectively, for $U$ posed in the set $Y(x)$.
The \underline{global} diffusion of $CO_2(g)$ is described at the macroscopic level $\Omega\times (0,T)$ via the macroscopic balance law:

\begin{equation}
\partial_t(\phi(x)u(x,t))+ div_x (-\mathbb{D}(x)\nabla u(x,t))+ \mathcal{E}(x,t) =\phi(x)f(x,t)
\end{equation}
Here we have:
\begin{itemize}
\item $\phi(x)$ - volumetric porosity of the material;

\item $\mathbb{D}(x)$ - effective diffusion tensor;

\item $\mathcal{E}(x,t)$ - exchange term modeling the flux of $CO_2$ escaping to/from the macroscopic domain $\Omega$.
\end{itemize}

Boundary conditions and initial conditions are needed to complete the macroscopic description. Note that without the exchange term $\mathcal{E}(\cdot)$ the same macroscopic equation for $CO_2(g)$  has been reported earlier in \cite{Florin}.

The precise structure of the volumetric porosity and of the effective diffusion tensor (or of the corresponding tortuosity tensor) are main objects of study in the context of the theory of porous media; see e.g. \cite{Bear} (for a view on the general theory) and 
\cite{Muntean_book} (for a more didactic view). Here we have the porosity $\phi(x)=\frac{|Y(x)|}{|\Omega|}$, while the precise structure of the effective tensor $\mathbb{D}(x,t)$  appears to be not so clear unless further information on the distribution of microstructures of the target material is available. One may even say that finding the correct explicit  calculation formulae for  the entries of $\mathbb{D}(x,t)$  is like finding the holy grail of porous media research. If one is ready to assume for instance that the porous media is a $2D$ rectangular block that has balled-shaped (sufficiently small) microstructures placed locally periodically with respect to the space variable $x\in\Omega$ and evolving in a smooth fashion\footnote{The growing or shrinking balls preserve their shape along the time evolution.} as time elapses, then one can obtain explicitly the structure of $\mathbb{D}(x,t)$ until neighboring balls touch each other -- an indicator of a possible occurrence of a local clogging situation; for this, see \cite{Adrian_SIAP}.

Similarly to the discussion of the macroscopic case, the \underline{local}, patch-wise, diffusion of $CO_2(aq)$ is described at the microscopic level $Y(x)\times (0,T)$ via:
\begin{equation}
\partial_t U(x,y,t)+ div_y (-D\nabla_y U(x,y,t))= F(x,y,t).
\end{equation}
The main issue in two-scale modeling procedure is to decide on how to couple the information among the scales. There are couple of ways to proceed. In this case, we benefit from what period homogenization techniques teach us. Consequently,  
for any $y\in\partial Y(x)=:\Gamma(x)$,  we consider the micro-macro flux boundary condition:
\begin{equation}\label{multiflux}
- D\nabla_y U(x,y,t)\cdot n_y =\frac{1}{\upsilon}\left(U(x,y,t)-H\phi(x,t)u(x,t)\right),
\end{equation}
that is, at each $x\in\Omega$, the microscopic fluxes are proportional to deviations from local equilibria as described by the Henry law (with Henry constant $H$). The characteristic time scale that is associated to this deviation from a local equilibrium situation is denoted here by $\upsilon$.  We take $H>0$ and $\upsilon>0$. The use of the Henry's law at the right-hand side of \eqref{multiflux} may seem a mathematically convenient choice, as it delivers a linear explicit expression of the flux term $- D\nabla_y U(x,y,t)\cdot n_y$ at $\Gamma(x)$. However, if in practice, a near equilibrium holds at gas-water interfaces, then this choice represents closely what happens physically. If strong deviations from local equilibrium occur, then most likely the right-hand side of \eqref{multiflux} does not reflect anymore the reality and needs to be changed, probably using a data-driven approach. As this relation connects intimately microscopic and macroscopic information, its general structure is largely unknown. This is precisely one of the important  places where inverse problem formulations can turn to be very useful, provided they are adapted to be applicable to such multiscale framework. 

Finally, the missing element is the structure of the averaged exchange  term $\mathcal{E}(x,t)$. This is here given by
\begin{equation}\label{E}
\mathcal{E}(x,t) :=
\frac{1}{|Y(x)|}\int_{\Gamma(x)} D\nabla_y U(x,y,t) \cdot n_yd\sigma_y.
\end{equation}
One can easily see that the choice \eqref{E} preserves the concentration of mass when crossing $\Gamma(x)$. The precise choice for the structure of $\mathcal{E}$ pointed out in \eqref{E} is not at all obvious. This conservative expression is borrowed from results obtained in the context of  periodic homogenization asymptotics applied to diffusion problems through high-contrast media.
The initial conditions $\phi(x)u(x,0)$ and $U(x,y,0)$ must also be described for all $x\in \bar \Omega, y\in \overline{Y(x)}$.
We observe that a simple application of Gau\ss's Theorem gives
\begin{equation}
\frac{\partial}{\partial t}\int_{Y(x)} U(x,y,t)dy = \int_{\Gamma(x)}D\nabla_y U(x,y,t)\cdot n_y d\sigma_y + \int_{Y(x)}Fdy.
\end{equation}

Using the choice of the micro-macro flux boundary conditions (\ref{multiflux}), we can represent the micro-concentration $U$ (via the corresponding Green function integrated over $\Gamma(x)$) as 
\begin{equation}\label{mNice}
U(x,y,t)=\int_0^t \mathcal{G}(x,t-\tau, H, Y_x,\phi(x))u(x,\tau)d\tau.
\end{equation}
Explicit expressions for $\mathcal{G}(\cdot)$ are available for simple shapes of microstructures like $\partial Y(x)=\partial Y=\Gamma$ with $Y$ being a rectangle or a disk. In general, only numerical approximations can give access to (\ref{mNice}). In either case, we are led to the following macroscopic evolution equation:
\begin{equation}
\partial_t\left(\phi u(x,t)+\int_0^t  \mathcal{G}(x,t-\tau, H, Y_x, \phi(x))u(x,\tau) d\tau\right)+ div_x (-\mathbb{D}\nabla u(x,t))=\phi f(x,t).
\end{equation}
The term 
$$\int_0^t  \mathcal{G}(x,t-\tau, H, Y_x, \phi(x))u(x,\tau)d\tau$$
models a storage effect with long or short memory (depending on the shape of the microstructure).

The source terms $f, F$ can be used to introduce production terms by chemical reaction in the discussion as it is of interest for carbonation scenarios. To reach the simplest possible two-scale carbonation scenario\footnote{We refer the reader to \cite{sebam_thesis} for a deeper discussion of the mathematics of this situation.}, one may take $f=0$ in $\Omega$ and $F$ to be proportional to the reaction rate corresponding to \eqref{carb} formulated at the level of $|Y(x)|$ for each $x\in\Omega$. 

It is worth noting at this stage that the two-scale model presented here is not the only possible way to describe coupled physical processes taking place on two different space levels. Prominent differences appear  for instance when modeling crystal growth over different spatial scales  (see \cite{ECK} as an example) or when micro-mechanics is coupled with macro-mechanics with or without the presence of phase transformations (see \cite{LIU} as an example). We do not elaborate further on the matter here.

\section{A brief literature review on multiscale carbonation models}\label{literature}

A nice recent review on modeling chemical reactions in porous media is Ref. \cite{Detmann}. Here the discussion refers to the macroscopic scale only and it heavily relies on concepts involving mixture theories. The advantage of such approach is that it may provide good ways to model couplings between the chemistry and the mechanics of materials. 
Staying away from modeling mechanical deformations due to chemical attack, we refer to a number of comprehensive studies on homogenized and two-scale carbonation models starting to appear cca. 15 years back focusing exclusively on the durability of concrete; see  e.g. \cite{Portugal,Adrian_Gakuto,sebam_thesis,Malte_thesis, mm_crm08} and their references. The rigorous derivation of homogenized carbonation models typically aimed at clarifying the structure of the constitutive laws entering already existing models. On the other hand, the introduction of two-scale carbonation models has open the possibility to address new questions particularly concerning the effect of the choice of microstructures on the overall macroscopically observed behavior. It is important to note
that the "two-scale" concept is not restricted to the prediction of carbonation, but can also be
useful when modelling, for instance, moisture transport, hydration, or sulphatation of the cementitious materials cf. e.g. \cite{Tasnim_thesis, Fatima2011261}. 

During the last years, sustained efforts have been paid in the applied analysis community to understand the basic qualitative properties of such models, particularly interesting topics being the large time behavior of the involved concentrations \cite{Aiki_large_sulfate, Kota} and their fast-reaction asymptotics  \cite{Evans}.
 
If sufficient regularity on the geometry of the microstructure, parameters, and respectively on initial and boundary data is available, then good concepts of solutions exist and such two-scale models tend to be well-posed in the sense of Hadamard. To get a flavor of the involved mathematical analysis arguments, we refer the reader to \cite{friedman_mm87} (classical solutions), \cite{Escher} (strong solutions), \cite{Maria_Adrian} (weak solutions), and to \cite{ineq} (other concepts of solution). The remaining prominent scientific challenges in this context are essentially related mainly to two aspects:
\begin{itemize}
\item[(I)] computability of the proposed multiscale models,
\item[(II)] validation of multiscale models against experimental data. 
\end{itemize} 
Two-scale computational schemes involving local representative unit-cells are usually
strongly related to the mathematical theory of homogenization \cite{Malte}. However, it turns out that approximation schemes can go beyond the settings usually handled by homogenization. For instance, convergent numerical approximation schemes for two-scale reaction-diffusion problems of the same type as the one described in the previous section were recently proposed in  \cite{Vladimir}  (two-scale finite differences) and in \cite{Omar_BIT}  (two-scale finite elements). With so much high performance computing power available these days, the aspect (I) is tractable at least in 2D. The 3D case is still computationally troublesome. The aspect (II) is more complex, and consequently, significantly less is done in this direction. Concerning the validation of carbonation models both single- and two-scale, we warn about the list of possible obstacles as mentioned in \cite{Zetem}. From a more mathematical perspective,  
it seems that there is a lot of space open for explorations in the direction of multiscale inverse problems. As for now, we are only aware of the work \cite{Martin_Adrian_Omar} where the authors treated the question of the inverse stability of the solution to a two-scale problem with respect to the coupling parameter $\upsilon$ that enters the  micro-macro flux boundary conditions (\ref{multiflux}).

\section{Two multiscale modeling tasks}\label{tasks}

As already anticipated in section \ref{sec-abstract}, a population of $CO_2$ molecules becomes the main character in the two-scale mathematical play described in section \ref{sec-abstract}. We now look into two classes of carbonation problems -- The first refers to a bulk carbonation situation (Scenario A), while the second one (Scenario B) is relevant for a thin layer placed at the boundary of the target object under investigation. The starting point of the discussion is the observation that for single scale carbonation models a lot is known, particularly what concerns materials involving the Portland cement. A standard reference for cement chemistry is the monograph \cite{Taylor}, while a relevant bibliographic hint for a macroscopic reaction engineering approach to the carbonation of concrete is  \cite{Papadakis}; see also the follow-up papers in the same direction like \cite{Meier_Muntean_Chem_Engn_Sci} and references cited therein. The focus of most all  papers mentioned up to now in this text is on modeling mathematically processes related to the carbonation of concrete.

\subsection{Scenario A: Carbonation of concrete}

The first scenario is concerned with a multiscale approach to the modelling of the carbonation reaction in cementitious materials, viz.  
\begin{equation}\label{carb}
CO_2(g\to aq) + Ca(OH)_2(s\to aq)\to CaCO_3 + H_2O,
\end{equation}
when one accounts for diffusion processes taking place on two different spatial scales. On the macroscopic scale, carbon dioxide is transported fast through the network of large capillary pores. On the microscopic scale, local slow transport and chemical reaction take place in the small pores of the $CSH$ gel phase. The processes that  take place at the microscopic scale are modelled now within a representative unit cell by assuming an idealised, regular microstructure of the cement paste. 

%\underline{3 macroscopic unknowns}: $u_1 = [CO_2(g)], u_2 = Ca (OH)_2(s), u_3 = Ca CO_3(s)$. 
%\\
%\underline{1 microscopic unknown}: $U= [CO_2(aq)]$
The modeling exercise assigned to Scenario A reads: Formulate a two-scale model to describe the evolution of the mass concentrations $u(x,t) = [CO_2(g)](x,t)$, $U(x,y,t)= [CO_2(aq)](x,y,t)$ and of the mass concentrations for $Ca (OH)_2(s)$ and $Ca CO_3(s)$, respectively. Mind that the carbonation of $Ca (OH)_2(s)$ takes place at a very different spatial scale compared to the carbonation of $CSH$, for instance. This exercise is expected to lead the reader to a basic, fairly complete, two-scale carbonation scenario. 

One can extend the resulting system  of equations by describing additionally what happens with the carbonation of the $CSH$
gel as well as with the carbonation of some phases probably with minor importance $KOH$, $NaOH$, $Mg(OH)_2$, $CAH$, $C_2S$, $C_3S$. In this case, 
one may assume as well that the carbonation of the components $CAH$, $C_2S$, $C_3S$ takes place at the same spatial level as the carbonation of $CSH$, while the carbonation of  $KOH$, $NaOH$, $Mg(OH)_2$ takes place at the same spatial level as the carbonation of $Ca (OH)_2$.  To be able to handle this part, one really needs to extend the carbonation reaction \eqref{carb} to a much larger chemical reaction mechanism involving all reactive components; see e.g. \cite{PeterMunteanMeieretal.2008} or use \cite{Taylor} or other specialized textbook on cement chemistry. Perhaps the introduction of three relevant separate length scales would come in handy. This would lead to a three-scale model for the carbonation problem.

\subsection{Scenario B: Carbonation of nanolimes}

The second carbonation scenario that we have in view aims at shedding light on the effect of the surface area of nanolimes (i.e. $Ca(OH)_2$ nanoparticles) on the overall carbonation kinetics.  The nanolime particles are supposed to be distributed within a thin film coated on the surface e.g. of a stone statue that needs special reinforcement. Trusting \cite{Navarro}, the surface area of a typical nanolime particle is a  parameter expected  to have a strong effect on the nucleation and growth  of metastable phases. We wonder: What mathematical model can potentially capture this effect? 

Specifically, as a result of the carbonation reaction, amorphous calcium carbonate (abbreviated usually as ACC) forms firstly. Then this polymorph product likes to evolves into a series of metastable phases
\begin{center}($\star$)\quad
vaterite $\to$ aragonite $\to$ calcite,
\end{center}
depending very much on a number of critical parameters like temperature, humidity as well as the surface area of the $Ca(OH)_2$ nanoparticles. We refer the reader, for instance,  to \cite{Navarro,buildings} for more information on this choice of scenario. Note that as Scenario  A is relevant for rather general concrete  carbonation scenarios, Scenario B is directly involved in heritage conservation.

The modeling exercise assigned to Scenario B reads: Formulate a two-scale model to describe the evolution of the mass concentrations 
\\
$u(x,t) = [CO_2(g)]$, $U_1(x,y,t)= [CO_2(aq)](x,y,t)$, $U_3(x,y,t)=[Aragonite](x,y,t)$, $U_4(x,y,t)=[Vaterite](x,y,t)$, $U_5(x,y,t)=[Calcite](x,y,t)$ 
(the last three unknowns being expected to describe the fact that $CaCO_3$ is polymorph).
This is a harder setup, in the sense that not only it is less explored in the literature but it offers at least two ways of tackling it. One way would be to consider a phase field model for the formation of the three phases, or to model the evolution of these phases explicitly, somehow similarly to the case of temperature-controlled formation of martensite, austenite, bainite as they arise during the casting of dual phase steels.   

\section{Discussion}

These notes should be seen as an addendum to the textbook \cite{Muntean_book}, where balance laws are now posed  in domains with distributed microstructures. They are addressed to the beginner researcher, and not to the expert in multiscale modeling and simulation. 

The literature review presented in section \ref{literature} is very much personalized, as it hints to what our research group and close collaborators did during the last years on the modeling, analysis and simulation of both  carbonation and sulphatation reactions in concrete.  Hence, it is not an exhaustive review as many relevant approaches on related topics are not mentioned.  

One of the main restrictions which make a two-scale modeling strategy possible is that the relevant physical processes take place at separated length scales. It would be interesting to see whether the discussion in section  \ref{sec-abstract} has a chance to be rewritten to cope with information coming from strongly interacting scales. 

Another aspect that is worth to be investigating from both modeling and mathematical analysis viewpoints would be to allow locations in $\Omega$, where the volume of the cell $Y(x)$ degenerates (i.e. either there is no wetness at all, or the wetness localizes on lower dimensional objects). It is likely that the methodology reported in \cite{singular} would be applicable to treat  at least partly such a case.

We do not provide answers to the modeling exercises presented in section \ref{tasks}, but we are happy to interact with the reader about possible partial solutions. The crucial hint is to adapt Showalter's methodology explained in section \ref{sec-abstract} to both Scenario A and Scenario B.

\section*{Acknowledgments} I am indebted to M. B\"ohm and S. A. Meier (both from Bremen, Germany) and to R. E. Showalter (Oregon, USA) for the things I have learned from them along the years  about modeling multiscale processes in terms of distributed microstructures.  I thank the editors of this volume for their efficient support as well as to the MACH initiative for promoting interactions between mathematicians and applied scientists within the field of cultural heritage. My work is partly supported by  the Swedish Research Council's project ``{\em  Homogenization and dimension reduction of thin heterogeneous layers}" (grant nr. VR 2018-03648). 

%\bibliographystyle{plain}    
%\bibliography{AM-refs-MACH} 

\end{document}